\documentclass[usenatbib,usegraphicx]{mn2e}

\begin{document}

\title{The SDSS DR7 Galaxy Angular Power Spectrum}
\author[B. Hayes et al.]{Brett Hayes$^1$, Robert Brunner$^1$, and Ashley Ross$^2$ \\ $^1$Department of Astronomy, University of Illinois, Urbana, IL 
\\ $^2$Institute of Cosmology \& Gravitation, University of Portsmouth, Portsmouth, UK}

\maketitle
\begin{abstract}
We calculate the angular power spectrum of galaxies selected from the Sloan Digital Sky Survey (SDSS) Data Release 7 (DR7) by using a quadratic estimation method 
with KL-compression. The primary data sample includes over 18 million galaxies covering more than 5,700 square degrees after 
masking areas with bright objects, reddening greater than 0.2 magnitudes, and seeing of more than 1.5 arcseconds.  We test for 
systematic effects by calculating the angular power spectrum by SDSS stripe and find that these measurements are minimally 
affected by seeing and reddening.  We calculate the angular power spectrum for $\ell \le 200$ multipoles by using 40 bandpowers for the full 
sample, and $\ell \le 1000$ multipoles using 50 bandpowers for individual stripes.  We also calculate the angular power spectrum for this sample
separated into 3 magnitude bins with mean redshifts of $z = 0.171$, $z = 0.217$, and $z = 0.261$ to examine the evolution of the angular 
power spectrum.  We determine the theoretical linear angular power spectrum by projecting the 3D power spectrum to two dimensions for 
a basic comparison to our observational results.  By minimizing the $\chi^2$ fit between these data and the theoretical linear angular power 
spectrum we measure a loosely-constrained fit of $\Omega_m = 0.31^{+0.18}_{-0.11}$ with a linear bias of $b = 0.94 \pm 0.04$.
\end{abstract}

\section{Introduction} 
\label{Introduction}

The angular power spectrum, $C_{\ell}$, is a statistical measure that quantitatively characterizes the large scale angular distribution of matter \citep{peebles73}.
Therefore, calculating the angular power spectrum of galaxies is useful as both a method of data compression, reducing clustering 
information of an arbitrary number of galaxy positions down to a set of $C_{\ell}$ and their corresponding window functions, 
and also since the $C_{\ell}$ values derived from the observations can be easily compared to theoretical predictions.

Calculations of angular power spectra are well known to cosmologists for their usefulness in studying the Cosmic Microwave 
Background (CMB), as the CMB provides a detailed and precise measurement of the density variations in the early 
universe (e.g., \citealt{smoot92,netterfield02,spergel07}).  However, to study large scale structure in other eras, it is necessary to 
analyze how mass clusters by using galaxies as a tracer of the underlying dark matter distribution. 

Angular power spectra of galaxies have been calculated for galaxy surveys with various depth and survey area 
(e.g., \citealt{huterer01,blake04,frith05}) including the SDSS (\citealt{tegmark02}, hereafter T02; \citealt{blake07,thomas10}).  
By using angular power spectra to calculate galaxy clustering, we study the Fourier modes of the galaxy distribution; this 
method is most sensitive to large scale effects. Recent galaxy surveys such as the APM Galaxy Survey \citep{maddox90}, the 
Two Micron All Sky Survey \citep{skrutskie06}, and the SDSS \citep{abazajian09} have cataloged large areas of the sky, 
thereby providing enormous numbers of galaxies for which we can measure angular clustering.  However, to date the galaxy
angular power spectrum has not been calculated for the full SDSS main galaxy sample.  In this paper, we address this deficiency.

The angular power spectrum is useful for large scale clustering, while it is complemented by the two-point angular correlation 
function on small scales.  The two-point angular correlation function (e.g., \citealt{brunner00,myers07,ross10}), which is related to 
the angular power spectrum by the Legendre transform (T02), is more sensitive to smaller scale clustering because the calculation is 
done in configuration space where the distances between nearby pairs of galaxies can be calculated faster.  This makes the two-point 
angular correlation function advantageous to use on scales where non-linear evolution is important.  This regime is also where the
angular power spectrum at large $\ell$ is more difficult to measure and model, partly due to correlations introduced between the $C_{\ell}$.

To calculate the angular power spectrum, we want to find the most probable parameters $C_{\ell}$ that could produce the data we 
observe.  To do this, we need the likelihood function of the angular power spectrum, which is proportional to the probability of 
the data given the $C_{\ell}$.  Though in theory we would like to know the entire likelihood function, calculating this 
$\ell_{max}$-dimensional function is difficult \citep{oh98}.  Fortunately, since we are only interested in the most probable $C_{\ell}$, we
only really need to know the maximum of this function.

To determine the $C_{\ell}$ that maximize the likelihood function, we use the quadratic estimation method (\citealt{tegmark97a,bond98}, hereafter BJK98).  This technique fits 
a quadratic function to the shape of the likelihood function for some initial angular power spectrum, finds the $C_{\ell}$ that maximize this 
quadratic, and uses these $C_{\ell}$ for a new quadratic fit to iteratively converge to the true maximum of the likelihood function.
Once we have found the angular power spectrum of galaxies, we can use the results to infer what cosmological 
parameters are consistent with the measurement (e.g., \citealt{jaffe99}).  

In this paper, we discuss the SDSS DR7 data, our selected sample and subsamples, and our systematic tests and masks in Section \ref{Data}.  
In Section \ref{Method}, we discuss our pixelization scheme, KL-compression,  and the quadratic angular power spectrum estimation method of BJK98 
in detail.  In Section \ref{Results}, we apply this estimator to the complete SDSS DR7, selected subsamples, and individual SDSS stripes, 
and present the results.  We construct a theoretical linear angular power spectrum to compare with the observational results, and we extract 
cosmological matter density and linear bias from this computation in Section \ref{Theory}.  Finally, we discuss our results in Section
\ref{Discussion}, and conclude the paper in Section \ref{Conclusion}.

\section{Data}
\label{Data}

The data for these measurements were taken from the SDSS Data Release 7, the final data release of SDSS-II.  The Sloan Digital Sky Survey \citep{abazajian09}
is a multi-filter imaging and spectroscopic survey using the 2.5 meter telescope at Apache Point Observatory 
that begun operation in 2000, and ended with the SDSS-II in 2008. The imaging observations are taken simultaneously in 5 filters (u, g, r, i, and z) as the 
telescope drift scans across the sky \citep{gunn98}.  The SDSS DR7 covers 11,663 square degrees in a striped fashion.  

The SDSS DR7 also provides photometric redshifts and redshift errors for each galaxy \citep{abazajian09}.  The SDSS has measured over 900,000 galaxy spectra
and uses these as a reference set to find the 100 nearest neighbors of a photometrically observed galaxy in color-color space.  The photometric redshift is 
estimated by fitting a hyperplane to these neighbors, and the error is determined by the mean deviations from the best-fit hyperplane \citep{csabai07}.  As
we require the galaxy redshift for analysis of our results, any
galaxy without both a photmetric redshift and associated error is not used in our calculation.  In SDSS DR7, the 
rms error of the photometric redshift estimation is 0.025, while for our samples it varies from 0.038 in the brightest sample to 0.064 in the dimmest.

\subsection{Area} 
\label{Area}

We begin by selecting a large, contiguous area of DR7, from stripes 9 to 37, an area of 7,646 square degrees before 
masking.  Each stripe is 2.5 degrees wide in eta (the survey latitude), and variable length in lambda (the survey longitude). 
Typically, however, the stripes are 100--120 degrees long.  Using this large area allows us to use a bandpower resolution of up to 4 
multipoles per bandpower when calculating the angular power spectrum for the full sample (see Section \ref{Selecting Bandpowers}).
Since this area is centered around the North Galactic Cap, we avoid the worst areas of reddening due to the Galactic disk.  
After masking for observational effects (e.g., reddening, seeing, bright stars; see Section \ref{Systematics}), our sample includes 
18.9 million galaxies over 5,763 square degrees of the SDSS Northern Galactic Cap ellipsoid.

\subsection{Systematics}
\label{Systematics}

\begin{figure}
  \includegraphics[width=0.5\textwidth]{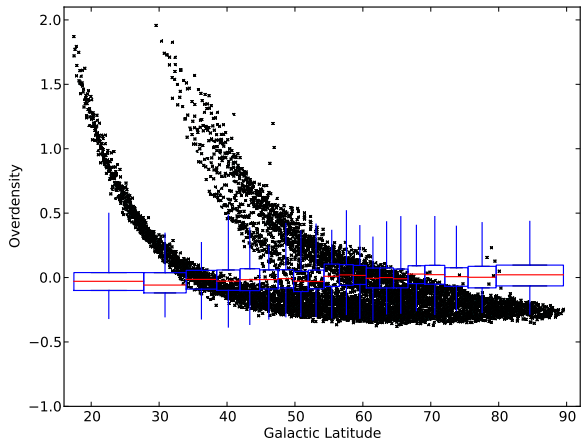}
  \caption{Points in black are the pixelized stellar overdensities, as a function of Galactic latitude at HEALPix resolution 64.  The exponential 
    falloff of the Galactic disk is seen here twice, at high Galactic latitude we see the falloff of the stars toward the Galactic center and at low
    Galactic latitude we see the stars in the direction of the Galactic anticenter.  We group the 
    pixelized galaxy overdensities by Galactic latitude into 20 bins, which are graphed as a box plot.  For each bin, the median galaxy 
    ovedensity is plotted in red, the end of the boxes mark the 25\% and 75\% quartiles, and the end of the whiskers mark the minimum and maximum 
    overdensities in that bin.}
  \label{Star Galaxy Overdensity}
\end{figure}

The data that we use span a wide range of Galactic latitudes, and we have considered the effect of stellar density on our galaxy samples.
Bright stars in our Galaxy could possibly obscure background galaxies \citep{ross11}, or faint stars could be misclassified as galaxies by the star-galaxy
separation routine.  To examine these possibilities, we have calculated the galaxy overdensity and stellar overdensity separately, applied our masks, and 
plotted these overdensities versus Galactic latitude in Figure \ref{Star Galaxy Overdensity}.  We see two exponential falloffs in the stellar overdensity
which correspond to the two edges of the SDSS dipping toward the Galactic disk, the high Galactic latitude exponential comes from the side of the 
SDSS in the general direction of the Galactic center and the low Galactic latitude exponential from the side near the Galactic anticenter, 
while the galaxy overdensity is consistent with zero at all Galactic latitudes in our sample.  For the large pixel sizes we use in 
the following calculations, obscuration by bright stars does not have a large effect on the galaxy overdensity, and at even at the lowest 
magnitude we use, star-galaxy separation is accurate at the 95\% confidence level \citep{lupton01} so we observe no effect on the galaxy 
overdensity from stars.

\begin{figure}
  \includegraphics[width=0.5\textwidth]{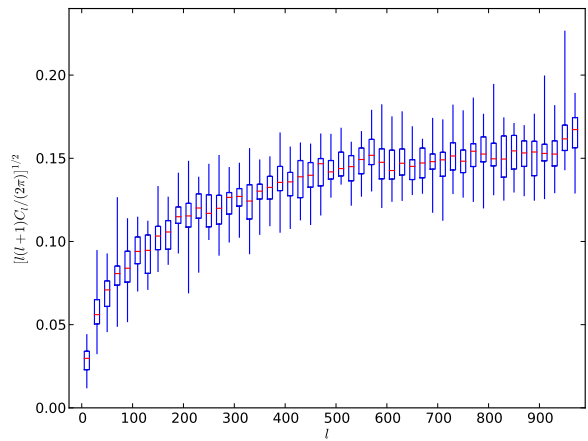}
  \caption{Box plot of the angular power spectra of galaxies with dereddened r-band magnitudes between 18 and 21 for the individual stripes 
    9 through 37. The median is in red, the 25\% and 75\% quartiles marked as the edge of the boxes, and the minimums and maximums marked at 
    the end of the whiskers.}
  \label{Stripe APS}
\end{figure}

To test the homogeneity and observational character of the data, we calculate the angular power spectrum separately for each stripe, using 
the method discussed in Section \ref{Method}.  If there is a significant deviation in the angular power spectrum from stripe to stripe, 
observational systematics might dominate over the real density variations of the combined stripe data that makes our full sample.  
To test for these systematics, we have calculated angular power spectra of each SDSS stripe from stripe 9 to stripe 37 after masking, 
with each of the $\mathcal{C}_{\ell}$ including an identical range of $\ell$.  The angular power spectra from each stripe
are remarkably consistent with each other, which is shown in the box-whisker plot in Figure \ref{Stripe APS}, and this shows that these 
observational systematics do not significantly alter the angular power spectra.
The only notable variation between stripes is that the edge stripes 9 and 37 have much larger error bars due to these stripes having the 
most pixels eliminated due to the seeing and reddening cuts.

We have also varied the seeing and reddening cuts to test their effects.  We have varied seeing cuts from 1.0 to 3.0 
arcseconds in 0.1 arcsecond intervals, and reddening cuts from 0.1 to 0.5 magnitudes in 0.05 magnitude intervals, but 
found that neither seeing nor reddening had a significant impact so long as a sufficient galaxy density remained to calculate 
the angular power spectra.  This is consistent with the cross correlations between galaxy density and reddening/seeing 
calculated by T02 for stripe 10.  Nevertheless, to minimize systematics in the SDSS galaxy sample, we have eliminated 
areas of seeing greater than 1.5 arcseconds and reddening worse than 0.2 magnitudes to be consistent with similar angular 
correlation function results \citep{ross07}, though others have used more stringent cuts \citep{wang12}.

\subsection{Subsamples} 
\label{Subsamples}

We have chosen our main sample to be from 18$^{th}$--21$^{st}$ magnitude in the extinction corrected r-band \citep{stoughton02}, 
with the faint limit chosen due to concerns about 
completeness in the sample past 21$^{st}$ magnitude.  Though the 95\% completeness r-band magnitude limit is 22.2 
\citep{abazajian09}, some galaxies at the fainter end of the 21-22 magnitude range are not detected or unusable due to large errors 
and we choose to limit our analysis to more complete samples.

We have chosen subsamples of our main sample for comparison to previous results, and to test for potential systematic errors on galaxy selection.
We first confirm our technique is consistent with the results from T02 up to 21$^{st}$ magnitude, so we have separated stripe 10 into
3 magnitude bins from 18--19, 19--20, and 20--21.  The comparison can be expected to be slightly different due to the use of the more 
complete DR7 data as opposed to the Early Data Release results that used galaxy probabilities (T02), in addition to the photometry 
calculation difference of magnitudes in SDSS data prior to DR2 \citep{abazajian04}.  We show these results in Section \ref{Results}.

We also measure the clustering attributes based on the brightness of the galaxies.  The apparently brighter galaxies 
cluster more strongly and are generally at lower redshift, thus we expect those to have more power in the angular power spectrum.
We create three new samples by separating the SDSS 
galaxies into 3 different r-band magnitude bins from magnitudes 18--19, 19--20, and 20--21. These magnitude ranges
are sufficiently bright to minimize the systematic effects of star-galaxy separation and variable sky brightness.
These samples have intrinsically different redshift distributions and luminosity functions, therefore the angular power spectra
of these samples will reflect these differences, and they are also useful as an important systematic test.

\subsection{Simulated Data Set}
\label{Simulated Data Set}

In addition to matching the published results from T02 and verifying that our results from all stripes across the SDSS DR7 
are consistent, we performed one additional test of the veracity of our quadratic angular power spectrum estimator.  We 
have generated simulated sky maps and compared the results from our quadratic estimator to the results from the 
HEALPix\footnote{See http://healpix.jpl.nasa.gov} angular power spectrum estimator anafast. We first generated a linear 
angular power spectrum as described in Section \ref{Theoretical Power Spectra} and used the HEALPix synfast routine to 
create ten pixelated sky maps at HEALPix resolution 2048. Second, we convert the pixel values in each of these ten sky 
maps to galaxy overdensities by using the average galaxy density of the SDSS DR7. Third,  we mask, in an identical manner 
to our treatment of the galaxy samples, each of these simulated full sky maps to the stripe 10 boundary as described in 
Section \ref{Pixel Masks}. Finally, we combine pixels to produce a degraded map with Healpix resolution 256. With these 
degraded sky maps, we calculate the angular power spectrum by using our quadratic estimator to these ten samples out to 
$\ell = 510$.

\begin{figure}
  \includegraphics[width=0.5\textwidth]{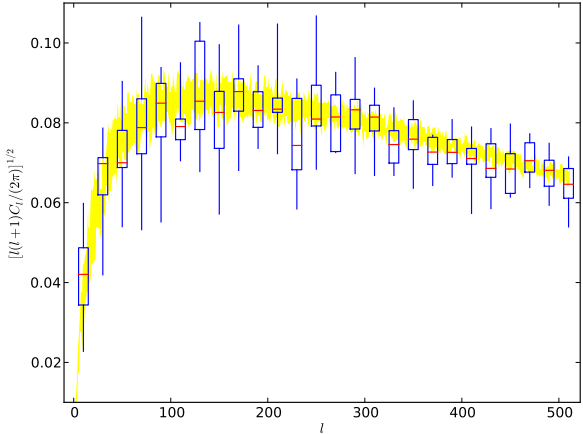}
  \caption{The results of our quadratic angular power spectrum estimation analysis of these 10 simulated maps is plotted 
    as a box plot with the median in red, 25\% and 75\% quartiles at the ends of the boxes, and the minimum and maximum 
    results at the ends of the whiskers. The yellow band shows the minimum and maximum angular power spectrum measurements 
    determined by the ten anafast measurements as described in the text.}
\label{Simulated APS}
\end{figure}

We also use synfast to generate the same ten maps at Healpix resolution 256, and calculate the angular power spectrum by 
using HEALPix angular power spectrum estimator anafast to provide a direct comparison to the results from our quadratic 
estimator. At resolution 256, we use the recommended $\ell=512$ for synfast and anafast, and performed a standard analysis
with anafast of the entire pixelated sky with no regression, masking, or cuts. We show these results along with the results 
from our quadratic estimator in Figure \ref{Simulated APS}. Both estimators show remarkable agreement, despite the fact 
that anafast is operating on a full sky map and our quadratic estimator is operating with the Stripe 10 window function. 
As a result, we feel our implementation of the quadratic estimator and the results we derive are robust.

\section{Method} 
\label{Method}

Angular power spectra attempt to measure the multipole moments, $\ell$, of a two dimensional distribution, in 
our case the galaxy density \citep{jaffe99}.  However, since photometric surveys only observe portions of the sky, all multipole 
moments cannot be individually determined \citep{tegmark96}; what is measured instead is a group of them simultaneously. Multipole 
moments are grouped into contiguous bands, called bandpowers, and we make the assumption that all moments
in the bandpower are equal (e.g., \citealt{huterer01}). The same computation is subsequently performed on the bandpowers as they would normally be on the 
individual multipole moments. This also serves to reduce the computation needed for the calculation \citep{borrill99}. First, 
we calculate the angular power spectrum by using the smallest bandpowers possible, and these bandpowers are averaged
together into larger bands to improve the signal-to-noise and reduce errors (BJK98).

Typically, Fourier methods are used to describe the distribution of a continuous population, but the galaxy distribution is 
discrete.  To calculate an angular power spectrum, we transform the discrete galaxy counts into
a continuous galaxy density distribution. To do this, the sky is divided into ``pixels'' and the galaxy
density in each pixel is calculated.  The calculation continues in the same way as it would with a CMB temperature
map (e.g., BJK98). Smaller pixels can tell us more information about the angular power spectrum, but the computation required
is highly dependent on the number of pixels \citep{tegmark97c}.

In this section, we first discuss how we pixelize and mask the data, followed by our selection of bandpowers in 
Section \ref{Selecting Bandpowers}.  In Section \ref{Calculating Bandpowers}, we extensively detail how we calculate
an angular power spectrum, beginning with KL-compression, the quadratic estimation technique, and the computational 
difficulty involved in this calculation.  In Section \ref{Interpreting Bandpowers}, we describe how these bandpowers
can be combined to produce higher signal-to-noise angular power spectrum estimates, and how to calculate
the window functions associated with these measurements.

\subsection{Pixelization} 
\label{Pixelization}

We have chosen to use a quadratic estimation approach to calculate the maximum likelihood  
of the angular power spectrum using KL-compression \citep{bond95,bunn95}. 
To force the discrete galaxy observations into a
continuous population, the sky is pixelated to determine the galaxy overdensity per pixel. 

We pixelate the sky using equal area pixels and remove areas that are outside the survey geometry, or have 
high seeing or reddening values.  Any pixels with less than 75\% usable area are not considered in the 
calculation.  In the end, the galaxy overdensity is calculated:

\begin{equation}
x_i \equiv \frac{G_i}{\overline{G}\Omega_i} - 1
\end{equation}
where $G_i$ is the galaxy count in pixel $i$, $\overline{G}$ is the average number of galaxies per square degree
over the survey area, and $\Omega_i$ is the area of the pixel in square degrees. Thus the data set of possibly
millions or more galaxies is reduced to a set of pixels that encodes the galaxy overdensities. The actual choice of 
pixelization technique, however, is important; and we have tested two different pixelization schemes, each with its own advantages.

\subsubsection{Pixelization Schemes} 
\label{Pixelization Schemes}

SDSSPix is a hierarchical, equal area pixelization scheme developed specifically for the SDSS by Max Tegmark, Yongzhong Xu, 
and Ryan Scranton\footnote{See http://dls.physics.ucdavis.edu/$\sim$scranton/SDSSPix/}. 
It uses the natural SDSS stripe geometry to divide the sky into pixels aligned with the SDSS survey coordinates, 
eta/lambda.  Pixels at a particular resolution have a constant width in eta, and a variable width in lambda to satisfy the 
equal area requirement.  While SDSSPix is useful because of the alignment of pixels with survey boundaries which makes seeing 
and reddening in pixels easier to quantify, the elongation of pixels away from the survey center interfered with the 
convergence properties of our algorithm described below.  This is because elongated pixels smooth density variations
preferentially in the direction of elongation while retaining that information in the perpendicular direction.  This 
increases the covariance between the smaller scale modes and drives increasing oscillation in high $\ell$ bandpowers with 
each iteration.

HEALPix is also a hierarchical, equal area pixelization scheme \citep{gorski05}, created for CMB experiments such as 
WMAP and Planck. It divides the sphere into 12 pixels at the base resolution, and higher resolutions recursively quarter these 
large pixels.  The benefit of using HEALPix is that while pixel boundaries have no relation to our observational data,
the pixels are not elongated as they are with SDSSPix.  Due to the stability of the quadratic estimation method using HEALPix, we have
opted to pixelize our data with HEALPix for our calculation.

\subsubsection{Pixel Masks}
\label{Pixel Masks}

\begin{figure}
  \includegraphics[width=0.45\textwidth]{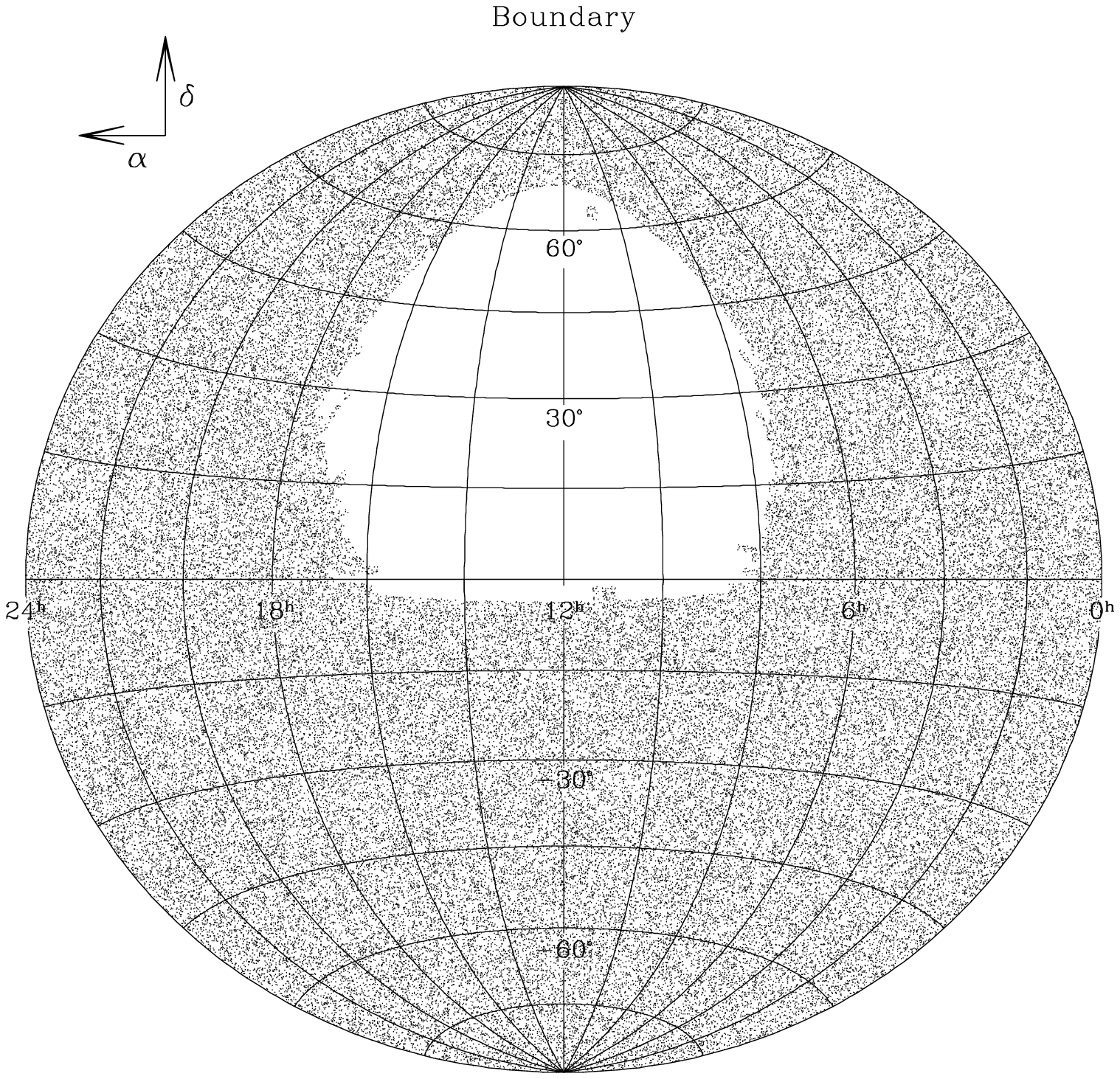}
  \includegraphics[width=0.45\textwidth]{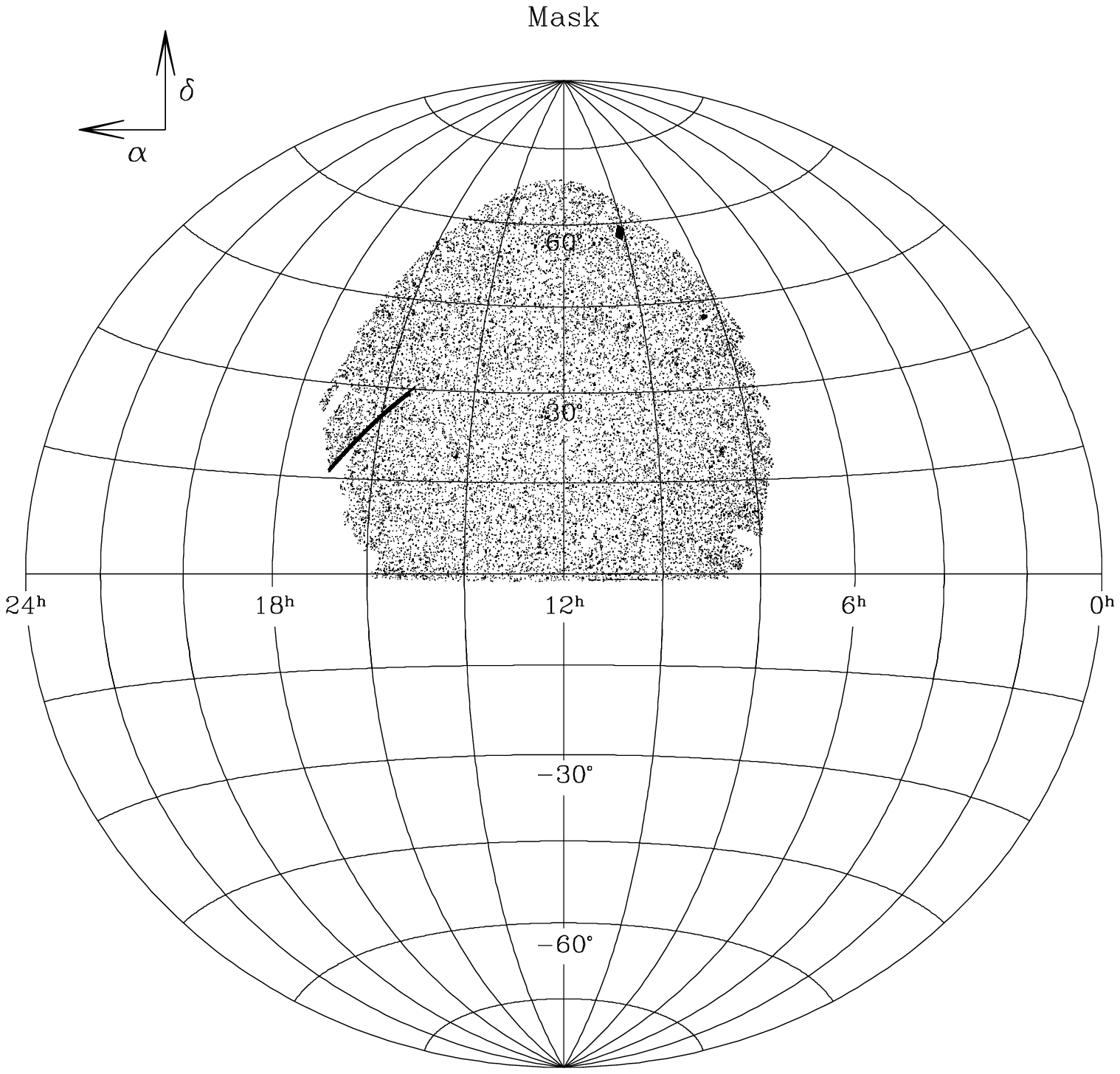}
  \caption{Top, the HEALPix pixels removed for being outside the chosen SDSS boundary, for clarity we have plotted a random sample of the masked pixels.  
    Bottom, the HEALPix pixels removed due to poor image quality.}
  \label{masks}
\end{figure}

\begin{figure}
  \includegraphics[width=0.45\textwidth]{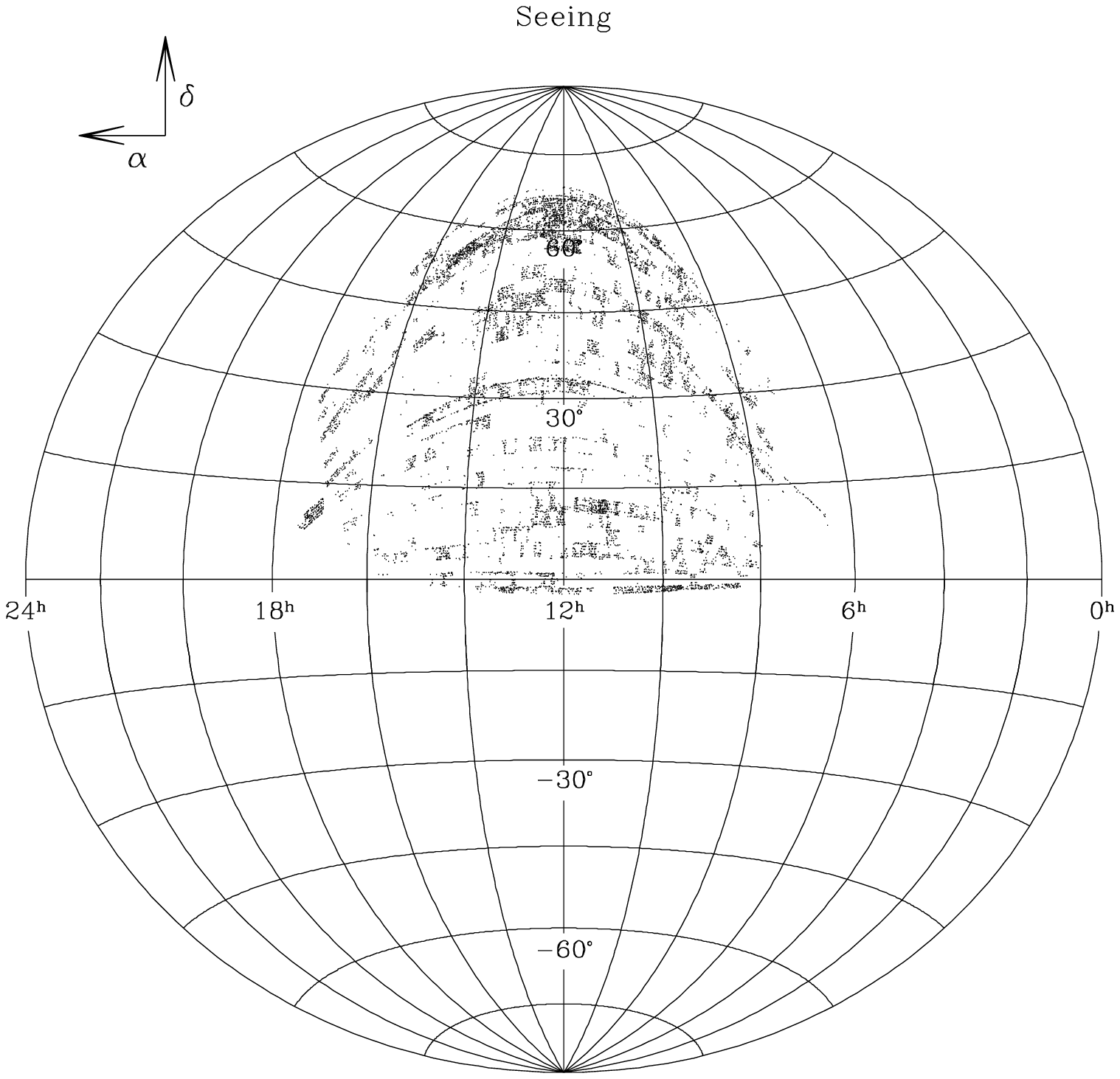}
  \includegraphics[width=0.45\textwidth]{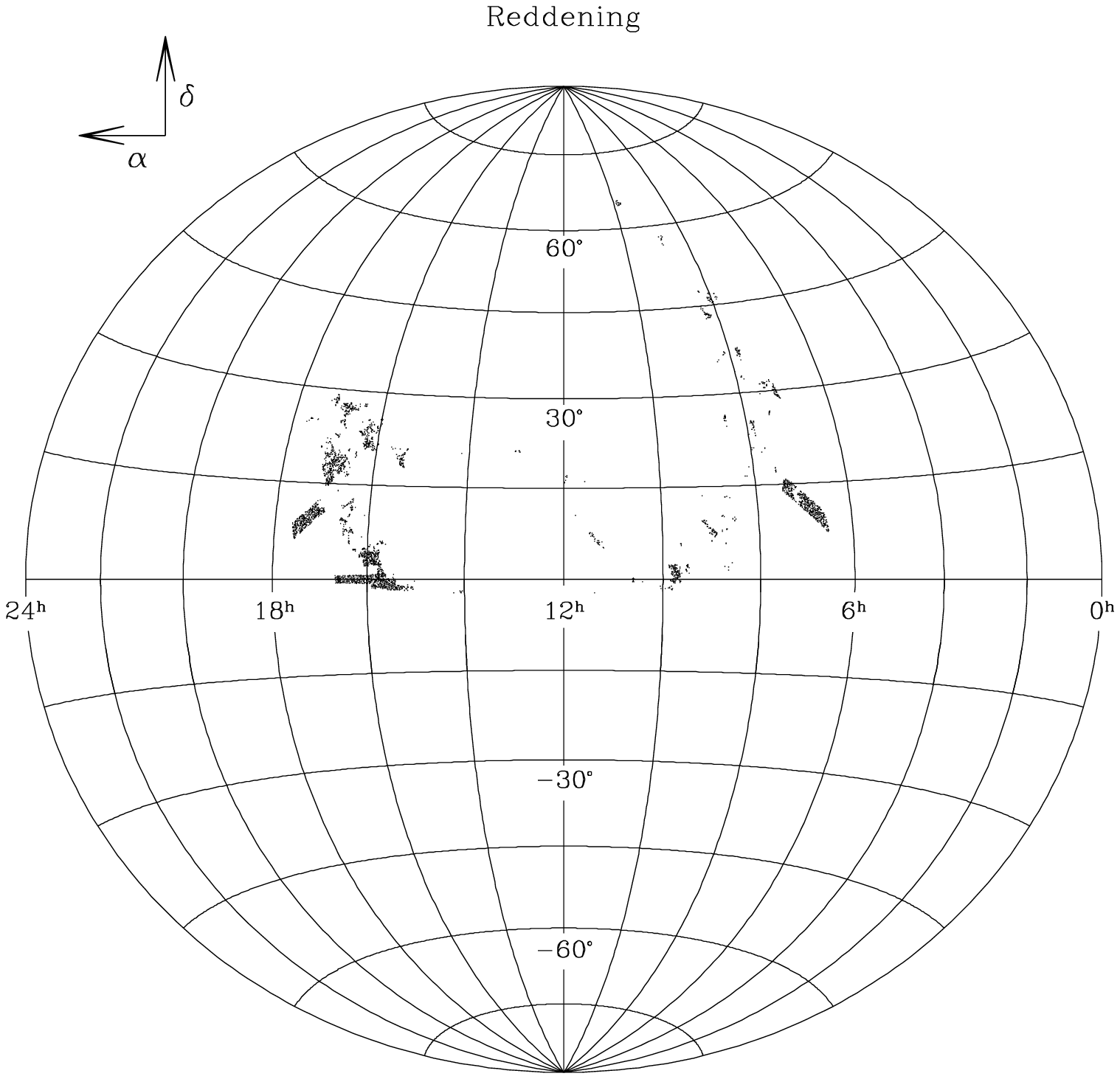}
  \caption{Top, HEALPix pixels removed for high seeing.  Bottom, HEALPix pixels removed due to high reddening.}
  \label{seeing}
\end{figure}

Masking with HEALPix is more complicated than with SDSSPix since pixels may overlap the survey boundaries.  For unbiased results, any
pixel that overlaps a boundary must not be considered in the calculation since it may have an unphysical overdensity.  Thus many pixels 
on stripe edges are masked.  We also eliminate pixels that are not contiguous with the primary SDSS observing footprint. A random sample 
of 100,000 of the pixels not used due to the boundary are shown in the top panel of Figure \ref{masks}, we have plotted only a sample 
to prevent obscuration of the coordinate lines. Furthermore, we must mask pixels due to areas with poor image quality, these pixels are 
also shown in the bottom of Figure \ref{masks}.  Additionally, we remove pixels where the mean seeing is more than 1.5 arcseconds, and 
pixels where the mean reddening is greater than 0.2, shown in the top and bottom panels of Figure \ref{seeing}, respectively.

\subsection{Selecting Bandpowers} 
\label{Selecting Bandpowers}

The first step in our approach is to select the initial fine bandpowers.  Multipole resolution is limited by $\Delta \ell \approx 
180^{\circ} / \phi$, where $\phi$ is the analyzed area's smallest angular dimension \citep{peebles80}. For this reason, we want the 
broadest survey possible.  Aside from being restricted to choosing bandpowers wider than this limit, the choice of the starting, 
ending, initial value, and widths of each bandpower is unrestrained, although some choices of initial values may cause 
non-convergence or singular matrices.  We chose initial bandpowers of equal widths, each $5 \ell$ wide for the full sample and 
$20 \ell$ wide for the individual stripes.  We use initial values based on a prior angular power spectrum; however, since the 
quadratic estimation method uses iteration, the final result is fairly insensitive to the input angular power spectrum.  We 
assume all $\mathcal{C}_{\ell}$ within a band to be constant \citep{huterer01}:

\begin{equation}
\mathcal{C}_\ell \equiv \frac{\ell(\ell+1)C_{\ell}}{2\pi} = \sum_{b} \chi_{b(\ell)} \mathcal{C}_b
\end{equation}
where $\chi_{b(\ell)} = 1$ while $\ell \in b$ and zero otherwise, and we define $\mathcal{C}_\ell$ according to standard convention 
\citep{bond00}.

We start with an initial fine binning, to determine where the power is inside the larger bandpowers that we later use. The Fisher 
information matrix (defined in Equation \ref{fisher}) is used to construct the bandpower window functions, and after we have 
performed the quadratic estimation to find the maximum likelihood, we will use these window functions to determine the correlation 
between bandpowers and individual multipole moments $\ell$.

\subsection{Calculating $\mathcal{C}_b$}
\label{Calculating Bandpowers}

Using only a knowledge of the survey geometry (or at least the region under consideration) and the 
assumed values for the bandpowers, we construct the covariance matrix \textbf{C}: 

\begin{equation}
C_{ij} \equiv \langle x_i x_j \rangle = \textbf{S} + \textbf{N}
\end{equation}
where, \textbf{S} is the signal matrix and \textbf{N} is the noise matrix. The 
assumed bandpower values $\mathcal{C}_b$ will only be approximate, which will make the covariance matrix approximate;
but this covariance matrix will be compared to the data and iteratively corrected to converge to the 
true bandpower values.  The signal matrix is 
calculated directly from the pixelated survey geometry using the assumed set of multipole values $\mathcal{C}_{\ell}$.
Using Legendre polynomials $P_{\ell}$ as the variance window functions, the calculated signal matrix \textbf{S} as shown by \cite{tegmark97c} is:

\begin{equation}
S_{ij} = \sum_{\ell} \frac{2\ell+1}{2\ell(\ell+1)} \mathcal{C}_\ell P_\ell (\cos \theta_{ij}) e^{-\ell(\ell+1)\tau^2} = \sum_{b} \mathcal{C}_b \textbf{P}_b .
\end{equation}
where $\theta_{ij}$ is the angle between pixels $i$ and $j$. 
The exponential factor is introduced to compensate for the smearing caused by a beam of width $\tau$. For pixels 
much larger than the beam, as is the case for a galaxy survey, this factor is negligible. The noise matrix, \textbf{N}, 
is modeled as a Gaussian random process and is diagonal \citep{huterer01}:

\begin{equation}
N_{ij} = \sigma_i^2 \delta_{ij} = \frac{1}{\overline{G}}\delta_{ij} ,
\end{equation}
where $\sigma_i$ is the rms noise in pixel $i$. 

\subsubsection{Karhunen-Lo\'{e}ve Compression}
\label{KL Compression}

Rather than perform the full calculation on the vector of overdensities $\textbf{x}$, we instead choose to 
transform into a signal to noise basis.  This is done by using KL-compression \citep{vogeley96,tegmark97b}.  
While this is often useful for data compression, with the high signal in our sample very few modes are discarded 
due to having greater noise than signal.  

We begin by solving the generalized eigenvalue equation:

\begin{equation}
\textbf{Sb}_{i} = \lambda_{i} \textbf{Nb}_i
\end{equation}
and normalizing such that $\textbf{b}^{T}_{i} \textbf{Nb}_{i} = 1$.  We reorder the vectors $\textbf{b}_{i}$
by the signal to noise ratio, $\lambda_{i}$, in descending order.  We discard modes with insufficient signal to noise, 
and we choose to keep those with $\lambda_{i} \ge 1$.  The remaining vectors $\textbf{b}_{i}$ form the columns of 
the matrix $\textbf{B}'$ that we use to transform the data vector $\textbf{x}' \equiv \textbf{B}'^{T} \textbf{x}$, 
as well as the signal, Legendre polynomial, and noise matrices $\textbf{S}' = \textbf{B}'^{T} \textbf{SB}'$, 
$\textbf{P}' = \textbf{B}'^{T} \textbf{PB}'$, and $\textbf{N}' = \textbf{B}'^{T} \textbf{NB}'$ (T02).

\subsubsection{Quadratic Estimation} 
\label{Quadratic Estimation}

From the new data vector $\textbf{x}'$, we perform the outer product to calculate the observed covariance matrix, 
$\textbf{x}' \textbf{x}'^T$, which will be compared to the constructed covariance matrix $\textbf{C}' = \textbf{S}' + \textbf{N}'$.

Now that we have a set of bandpowers that we want to determine, we calculate the $\mathcal{C}_b$ that have the highest 
probability of creating the observed data. A complete calculation of the likelihood function, although slow, is possible, 
but a local maximum can be found by using iteration with the following estimator (BJK98):

\begin{equation}
\label{estimator}
\delta \mathcal{C}_b = \frac{1}{2} (\textbf{F}^{-1/2})_{b b'}\ \textrm{Tr}\left[(\textbf{x}' \textbf{x}'^T-\textbf{N}')(\textbf{C}'^{-1} 
\textbf{P}'_{b'} \textbf{C}'^{-1})\right]
\end{equation}
where, the Fisher information matrix \textbf{F} is defined as:

\begin{equation}
\label{fisher}
F_{b b'} = \frac{1}{2}\ \textrm{Tr}\left(\textbf{C}'^{-1} \textbf{P}'_{b} \textbf{C}'^{-1} \textbf{P}'_{b'}\right)
\end{equation}

Equation \ref{fisher} provides the mechanism by which we can compare the covariance matrix obtained from the data 
$\textbf{x}' \textbf{x}'^T$ with the constructed covariance matrix $\textbf{C}'$. What this equation accomplishes 
is retrieving the $\mathcal{C}_b$ that produce a covariance matrix $\textbf{C}'$ that is identical to 
$\textbf{x}' \textbf{x}'^T$.  Note that we use $\textbf{F}^{-1/2}$ in Equation \ref{estimator} as advocated by 
\cite{tegmark98} for uncorrelated error bars and well behaved window functions.

By making an initial estimate of $\mathcal{C}_b$, and iteratively applying this equation, the estimator quickly converges
on a maximally probable set of bandpower values.  The error in bandpower $b$, given by $\sigma_{b} = \sqrt{(\textbf{F}^{-1})_{bb}}$,
is the smallest error any estimator can measure while estimating parameters from the sample itself due to the Cramer-Rao 
inequality \citep{kenney51,tegmark97a}.

\subsubsection{Computational Requirements} 
\label{Computational Requirements}

The quadratic estimation method is computationally complex, due to both a large amount of calculation required for matrix 
operations as well as large memory requirements to store these matrices.  We must consider computational feasability when 
making choices about the extent of the data that we will analyze.  At the scales of interest, we have found the processing 
time for a single processor scales as:

\begin{equation}
T \approx 6\ \textrm{days}\ \left(\frac{n_b}{40}\right)\ \left(\frac{n_i}{3}\right)\ \left(\frac{n_p}{6836}\right)^3
\end{equation}
and the memory requirements scale as:

\begin{equation}
M \approx 60\ \textrm{GB}\ \left(\frac{n_b}{40}\right)\ \left(\frac{n_p}{6836}\right)^2
\end{equation}
where $n_b$, $n_i$, and $n_p$ are the number of bandpowers, iterations, and pixels respectively. 
Typically only a few iterations are necessary; we allow 3 iterations to achieve convergence.  These are obviously highly 
dependent on the number of pixels $n_p$, and processing time and memory requirements become prohibitive much beyond
$10^4$ pixels \citep{borrill99}.  
As a result, we have made use of the National Center for Supercomputing Applications' (NCSA) 1,024 processor SGI Altix (Cobalt),
its successor the 1,536 processor SGI Altix (Ember), as well as the Pittsburg Supercomputing Center's 768 core SGI Altix (Pople) 
and 4,096 core SGI UV 1000 (Blacklight) for these calculations. 

\subsection{Interpreting $\mathcal{C}_b$}
\label{Interpreting Bandpowers}

\subsubsection{Averaging $\mathcal{C}_b$} 
\label{Averaging Bandpowers}

After defining the bandpowers and calculating the $\mathcal{C}_b$, we use the Fisher Information matrix to 
determine the correlation between bandpowers \citep{knox98}.  Narrow bandpower
window functions are preferred so that the error in one band measurement minimally affects other bands.

Though the Fisher matrix and $\mathcal{C}_b$ have already been calculated for the choice of bandpowers, we want 
to have a method of combining bandpowers to improve the signal-to-noise without recalculating 
using the computationally demanding quadratic estimator method.  For this we use the BJK98 method.

First, smaller bandpowers $b$ are averaged together into larger bandpowers $B$ (not to be confused with the KL-compression matrix 
$\textbf{B}$ defined earlier) using Equation \ref{Bandpower Average}.  We can combine any number of adjacent bandpowers to improve 
signal-to-noise, though combining bandpowers from sections of the angular power spectrum with significant structure will result 
in a loss of resolution in the areas of interest (BJK98).  

\begin{equation}
\mathcal{C}_B = \frac{\sum_{b \in B} \sum_{b'\in B'} \mathcal{C}_b F_{bb'}}{\sum_{b \in B} \sum_{b' \in B'} F_{bb'}}
\label{Bandpower Average}
\end{equation}

\begin{equation}
F_{BB'} = \sum_{b \in B} \sum_{b' \in B'} F_{bb'}
\end{equation}

The averaged Fisher matrix must be calculated to determine the errors on 
$\mathcal{C}_B$, which are $\sigma_{B} = \sqrt{(\textbf{F}^{-1})_{BB}}$ \citep{tegmark97c}.

\subsubsection{Calculating Window Functions}
\label{Plotting}

To represent the angular power spectrum visually, the data points are characterized not only by the values and errors,
but also by the width and position of the bandpowers they represent. The bandpower window functions are given by (T02):

\begin{equation}
\textbf{W} = \textbf{DF}^{1/2}
\label{Window Function}
\end{equation}
where $\textbf{D}$ is the diagonal matrix that makes the rows of $\textbf{W}$ sum to unity.  
The midpoints of the bandpowers, $\ell_{eff}$, can also be calculated.  Algorithmically, $\ell_{eff}$ is where half the 
power in the band comes from below and half from above that multipole (BJK98):

\begin{equation}
f_{Bb} = \sum_{b \in B} F_{bb'}
\end{equation}

\begin{equation}
\ell_{eff} = \frac{\sum_{b \in B} \ell f_{Bb}}{\sum_{b \in B} f_{Bb}}
\end{equation}

We calculate the filter $f_{Bb}$ while doing the averaging in Section \ref{Averaging Bandpowers}.  This filter 
function tells us how the power in larger bands is related to the power in the component smaller bands, and gives us 
information about how the power is distributed within the new larger bands (BJK98).  The edges of the band, $\ell^-$ 
and $\ell^+$, are defined to be where $\ell f_{Bb}$ drops to $e^{-1/2}$ of the peak power, and we plot these 
as horizontal error bars. The angular power spectrum at $\ell_{eff}$ can be plotted with horizontal error bars ranging 
from $\ell^-$ to $\ell^+$, with value $\mathcal{C}_B$ and vertical error bars $\pm \sqrt{(\textbf{F}^{-1})_{BB}}$.

\section{The SDSS Angular Power Spectrum} 
\label{Results}

\begin{figure}
  \includegraphics[width=0.45\textwidth]{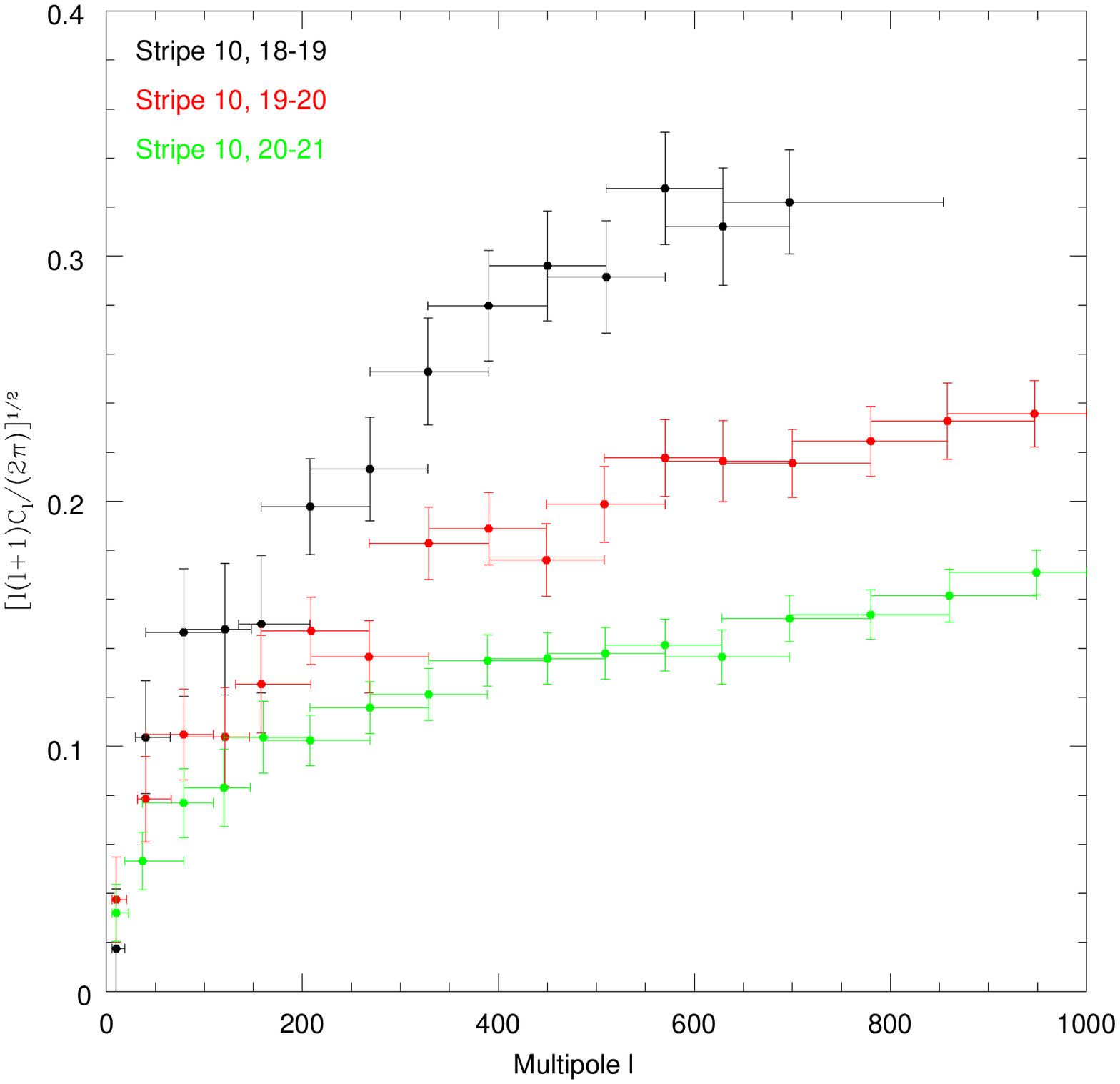}
  \includegraphics[width=0.45\textwidth]{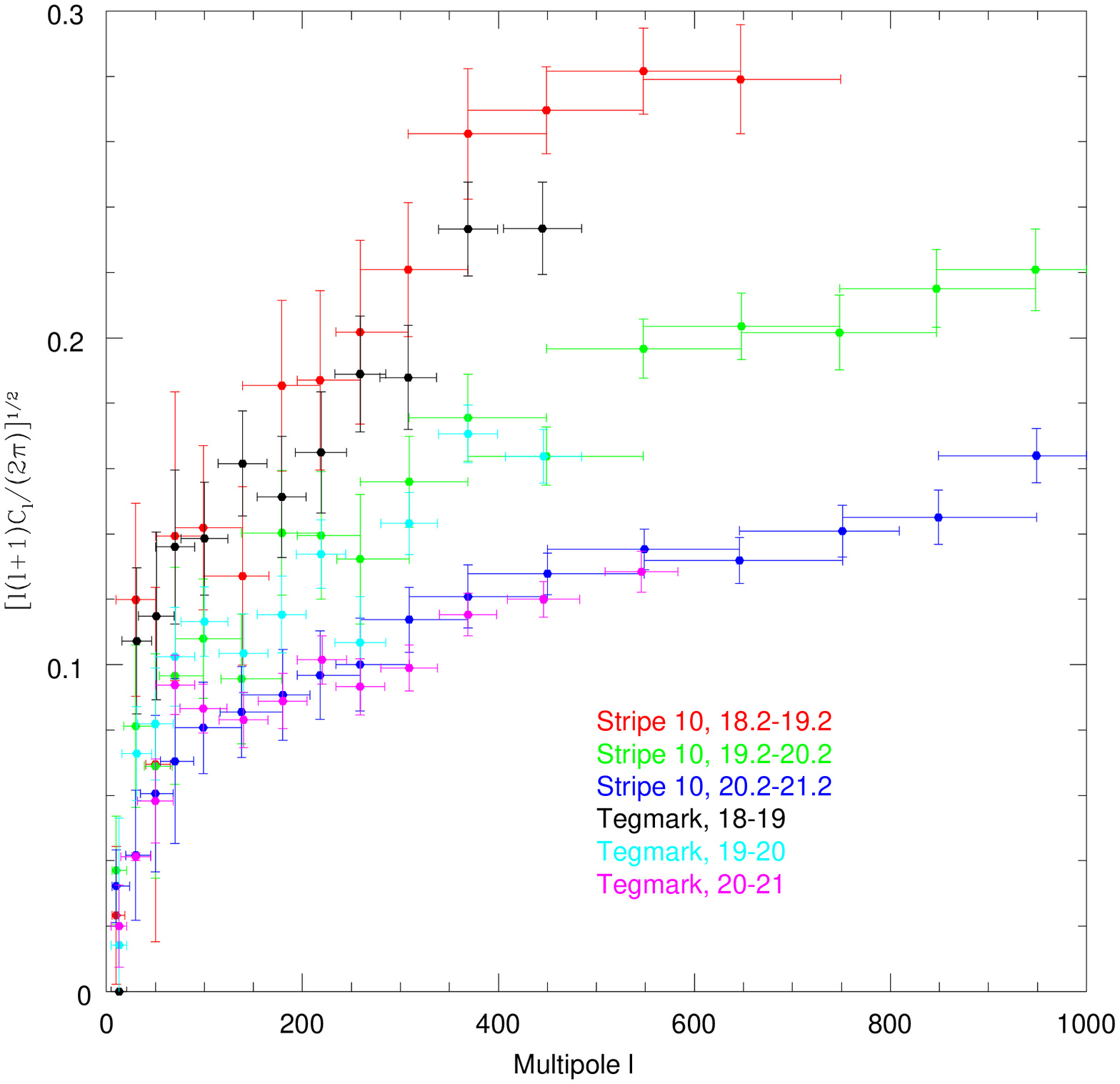}
  \caption{The top panel shows the angular power spectra of the 3 magnitude cuts on stripe 10.  The bottom panel shows 
    the magnitude shifted angular power spectrum in comparison with the results of T02.}
  \label{APS 1}
\end{figure}

The results of our angular power spectrum calculation for stripe 10 for $\ell < 1000$ are shown in the top panel of Figure 
\ref{APS 1}, separated by magnitude.  Though our results are consistently higher than those in T02 in all samples, we find 
that our results are still in agreement.  This is due to a known magnitude calculation error in early SDSS data, which 
miscalculated galaxy model magnitudes by roughly 0.2 mag \citep{abazajian04}.  When we shift the samples by 0.2 magnitudes 
to account for this difference, our results match very well with the previous results, typically within one standard deviation 
as shown in the bottom panel of Figure \ref{APS 1}.  Additionally, as we are using DR7 instead of the EDR, galaxy counts versus 
galaxy probabilities, and HEALPix rather than SDSSPix, we do not expect the results to exactly coincide.

\begin{figure}
  \includegraphics[width=0.45\textwidth]{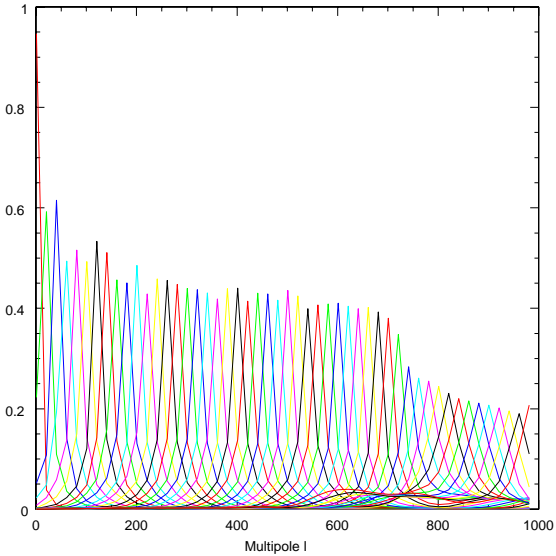}
  \includegraphics[width=0.45\textwidth]{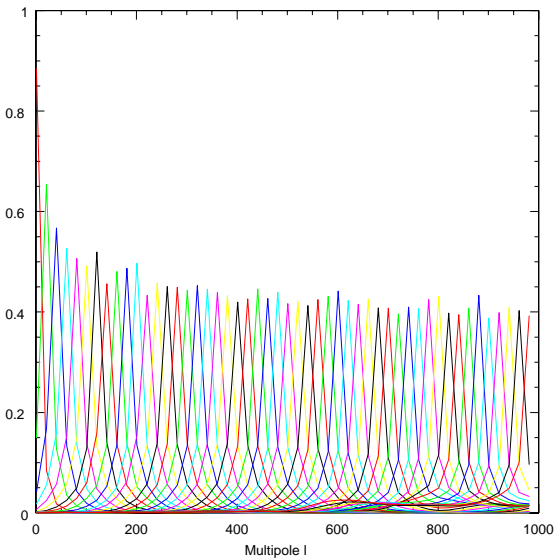}
  \caption{Window Functions - The window functions of each of the 50 bands, for the 18--19th bin (top) and 20--21st 
    magnitude bin (bottom) of stripe 10.}
  \label{Stripe 10 Window Functions}
\end{figure}

In addition, we not only need to know the final $C_{\ell}$, but to completely characterize the errors and the structure 
of each bandpower, we need to know the window functions.  The variance and covariance of the $C_{\ell}$ 
are derived from the Fisher matrix, and the bandpower window functions show what $\ell$ the power in a band 
comes from, so we prefer bandpower window functions to be as narrow as possible.  For illustration and comparison to T02, 
the bandpower window functions for the 18--19 magnitude bin of stripe 10 are shown in the top panel of Figure 
\ref{Stripe 10 Window Functions}.  We see that at about $\ell \sim 750$, the window functions 
become wider signifying that our signal has dropped below shot noise fluctuations, so bands beyond that are not used.  
In the other magnitude bins, our signal does not drop below shot noise fluctuations out to $\ell = 1000$ such as in the 
bottom panel of Figure \ref{Stripe 10 Window Functions}.  The window functions for other stripes are similar, and we 
have made these available online\footnote{All results discussed in this paper are available at 
$http$:$//lcdm.astro.illinois.edu/research/aps.html$}.

\begin{figure}
  \includegraphics[width=0.45\textwidth]{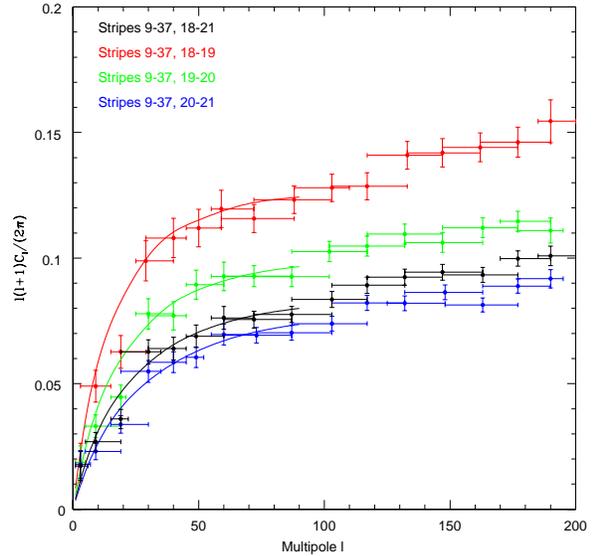}
  \caption{Angular Power Spectra - The spectrum of stripes 9 to 37, magnitudes 18--21 in black, 18--19 in red, 19--20 
    in green, and 20--21 in blue.  The solid lines are the best-fit theoretical linear power spectrum for $\ell < 90$.}
  \label{APS 2}
\end{figure}

The results of the angular power spectrum of our entire sample for $\ell < 200$, as well as for our magnitude separated 
subsamples, are summarized in Figure \ref{APS 2} and in Table 1.
The brightest and on average closest galaxies in the 18--19 r-band magnitude bin are the most highly clustered at all 
$\ell$ as expected. Below that is the 19--20 magnitude bin, and the least clustered at all $\ell$ is the 20--21 magnitude 
bin.  Also plotted are the linear theoretical angular power spectra discussed in Section \ref{Theoretical Power Spectra} 
for $\ell < 90$.

\begin{table*}
  \label{SDSS APS Table}
  \begin{center}
    \begin{tabular}{| cc | cc | cc | cc |}
      \hline
      Sample & 18--21 & & 18--19 & & 19--20 & & 20--21 \\
      \hline
      $\ell_{eff}$ & $\mathcal{C}_{B} \pm \sigma_{B}$ & $\ell_{eff}$ & $\mathcal{C}_{B} \pm \sigma_{B}$ & $\ell_{eff}$ & $\mathcal{C}_{B} \pm \sigma_{B}$ & $\ell_{eff}$ & $\mathcal{C}_{B} \pm \sigma_{B}$ \\
      \hline
      3 & 0.00030 $\pm$ 0.00020 & 3 & 0.00035 $\pm$ 0.00029 & 3 & 0.00035 $\pm$ 0.00025 & 3 & 0.00032 $\pm$ 0.00020 \\
      9 & 0.00072 $\pm$ 0.00020 & 9 & 0.00240 $\pm$ 0.00062 & 9 & 0.00110 $\pm$ 0.00030 & 9 & 0.00053 $\pm$ 0.00015 \\
      19 & 0.00129 $\pm$ 0.00028 & 19 & 0.00393 $\pm$ 0.00082 & 19 & 0.00199 $\pm$ 0.00043 & 19 & 0.00114 $\pm$ 0.00024 \\
      30 & 0.00393 $\pm$ 0.00061 & 29 & 0.00977 $\pm$ 0.00158 & 30 & 0.00605 $\pm$ 0.00094 & 30 & 0.00301 $\pm$ 0.00047 \\
      40 & 0.00409 $\pm$ 0.00058 & 40 & 0.01167 $\pm$ 0.00169 & 40 & 0.00593 $\pm$ 0.00087 & 40 & 0.00343 $\pm$ 0.00048 \\
      49 & 0.00475 $\pm$ 0.00062 & 50 & 0.01254 $\pm$ 0.00168 & 49 & 0.00800 $\pm$ 0.00104 & 49 & 0.00366 $\pm$ 0.00049 \\
      60 & 0.00581 $\pm$ 0.00069 & 59 & 0.01430 $\pm$ 0.00178 & 60 & 0.00860 $\pm$ 0.00105 & 60 & 0.00484 $\pm$ 0.00058 \\
      72 & 0.00571 $\pm$ 0.00051 & 72 & 0.01339 $\pm$ 0.00129 & 72 & 0.00861 $\pm$ 0.00078 & 73 & 0.00479 $\pm$ 0.00043 \\
      87 & 0.00601 $\pm$ 0.00050 & 88 & 0.01519 $\pm$ 0.00136 & 87 & 0.00858 $\pm$ 0.00074 & 87 & 0.00494 $\pm$ 0.00042 \\
      103 & 0.00698 $\pm$ 0.00054 & 103 & 0.01638 $\pm$ 0.00139 & 102 & 0.01053 $\pm$ 0.00084 & 103 & 0.00546 $\pm$ 0.00043 \\
      117 & 0.00795 $\pm$ 0.00057 & 117 & 0.01653 $\pm$ 0.00139 & 117 & 0.01098 $\pm$ 0.00084 & 117 & 0.00675 $\pm$ 0.00049 \\
      132 & 0.00853 $\pm$ 0.00059 & 133 & 0.01987 $\pm$ 0.00156 & 132 & 0.01201 $\pm$ 0.00088 & 132 & 0.00673 $\pm$ 0.00048 \\
      147 & 0.00891 $\pm$ 0.00059 & 147 & 0.02014 $\pm$ 0.00159 & 147 & 0.01126 $\pm$ 0.00083 & 148 & 0.00746 $\pm$ 0.00051 \\
      163 & 0.00871 $\pm$ 0.00056 & 162 & 0.02076 $\pm$ 0.00166 & 163 & 0.01256 $\pm$ 0.00088 & 163 & 0.00662 $\pm$ 0.00045 \\
      177 & 0.00997 $\pm$ 0.00061 & 177 & 0.02135 $\pm$ 0.00175 & 177 & 0.01314 $\pm$ 0.00090 & 177 & 0.00789 $\pm$ 0.00050 \\
      190 & 0.01018 $\pm$ 0.00078 & 190 & 0.02387 $\pm$ 0.00262 & 190 & 0.01231 $\pm$ 0.00111 & 190 & 0.00843 $\pm$ 0.00070 \\
      \hline
    \end{tabular}
    \caption{The SDSS Angular Power Spectrum for our entire sample and each of the 3 magnitude subsamples.  $\ell_{eff}$ is the 
      point in the band where half the power is from $\ell < \ell_{eff}$ and half the power is from $\ell > \ell_{eff}$, not 
      necessarily the center of the band.}
  \end{center}
\end{table*}

\begin{figure}
  \includegraphics[width=0.45\textwidth]{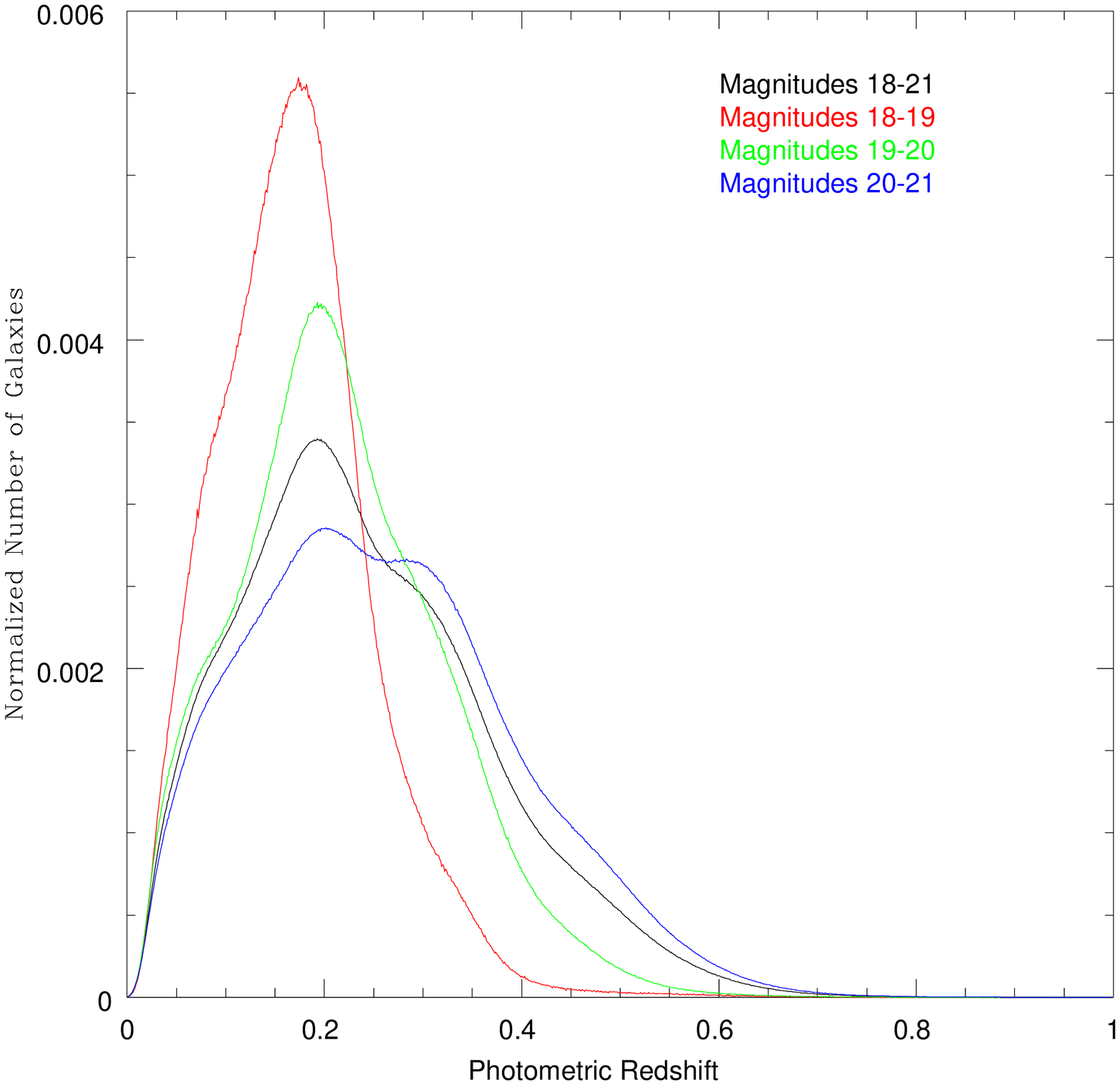}
  \caption{The normalized photometric redshift distribution of all galaxies in stripes 9 to 37, from 
    magnitude 18--21 in black, magnitude 18--19 in red, 19--20 in green, and 20--21 in blue.}
  \label{Zdist}
\end{figure}

\section{Theory}
\label{Theory}

\subsection{Theoretical Power Spectra} 
\label{Theoretical Power Spectra}

The statistical characterizations of galaxy clustering provided by our angular power spectrum measurements 
are only the first step. In order to constrain models of structure formation, we must compare these results to 
theoretical linear angular power spectra. To obtain theoretical $\mathcal{C}_{\ell}^T$, we project the  
linear 3D power spectrum $P(k)$, modeled with the fitting formulae of \cite{eisenstein98}, down to two dimensions. With $P(k)$, we can calculate 
the $\mathcal{C}_{\ell}^T$ we expect from a given theory (e.g., \citealt{huterer01}).  From \cite{crocce10} we
have the exact calculation for the theoretical linear angular power spectrum:

\begin{equation}
\mathcal{C}_{\ell}^T = \ell(\ell+1)/\pi^2 \int k^2 P(k) \Phi_{\ell}(k)^2\ dk
\end{equation}
where:

\begin{equation}
\Phi_{\ell}(k) = \int \phi(z) D(z) j_{\ell}(kr(z))\ b\ dz
\end{equation}

\begin{equation}
\phi(z) = \frac{1}{\overline{G}}\ \frac{d\overline{G}}{dz}
\end{equation}
where $D(z)$ is the growth function \citep{carroll92} and $j_l(kr)$ are Bessel functions, $b$ is the bias, and $r$ and 
$\overline{g}$ are the comoving distance and number density respectively. 
This simplifies if we use Limber's approximation \citep{limber53} to simplify the calculation of the Bessel functions:

\begin{equation}
\mathcal{C}_{\ell}^T \approx \frac{2\pi}{\ell(\ell+1)}\int \phi^2(z) D^2(z) P(\frac{\ell + 1/2}{r(z)}) \frac{H(z)}{r^2(z)} b^2\ dz\ 
\end{equation}

The theoretical power spectrum depends only on cosmological parameters through the 3D power spectrum and the bias, so 
we can use this dependence to infer constraints on these values.  The only knowledge it requires about the sample is the 
redshift distribution.  We calculate the redshift distribution by assuming the redshift of each galaxy is 
distributed as a Gaussian with mean equal to the observed photometric redshift and standard deviation equal to the error 
of the photometric redshift. We sample the distribution of each galaxy and then weight by volume and luminosity function 
constraints as in \cite{ross10} with the luminosity function of \cite{montero-dorta09}.

In Figure \ref{Zdist}, we show the photometric redshift distribution of our main sample of over 18 million galaxies, 
separated into photometric redshift bins of width 0.001 with $0.0 \le z < 1.0$.  We see that the peak of the sample is
at $z \sim 0.2$ and falls off rapidly past $z \sim 0.3$.  The redshift distribution is important because we must use 
it when we project the 3D power spectra to compare to our angular power spectra.  Also in Figure \ref{Zdist}, 
we have separated the 
redshift distribution into magnitude bins, and see the variations of photometric redshift distributions by magnitude, 
with the brighter bins being on average closer than the fainter bins.  The average redshifts of these samples are
$z = 0.171$ for the 18--19 magnitude bin, $z = 0.217$ for 19--20, $z = 0.261$ for 20--21, and $z = 0.243$ for the 
entire sample.

\subsection{Fitting Theory to Data} 
\label{Fitting Theory to Data}

To constrain cosmological parameters, we use a $\chi^2$ fitting technique to determine the calculated theoretical linear angular power 
spectrum that best fits the observed bandpower measurements \citep{tegmark97a}.  First, an average over the chosen bandpowers of 
the newly calculated $\mathcal{C}_\ell^T$ is made so that these can be compared \citep{knox99}:

\begin{equation}
\langle \mathcal{C}_B^T \rangle = \sum_{B'} W_{B B'}\ \mathcal{C}_{B'}^T
\end{equation} 
with the bandpower window function $W_{B B'}$ from Equation \ref{Window Function}.
We evaluate the following $\chi^2$ where $\textbf{F}$ is the Fisher matrix and $a_p$ are the cosmological parameters \citep{bond00}:

\begin{equation}
\chi^2(a_p) = \sum_{BB'} (\ln \mathcal{C}_B - \ln \mathcal{C}_B^T)\ \mathcal{C}_B F_{BB'} \mathcal{C}_{B'}\ (\ln \mathcal{C}_{B'} - \ln \mathcal{C}_{B'}^T)
\end{equation}

\begin{figure}
  \begin{center}
    \includegraphics[width=0.45\textwidth]{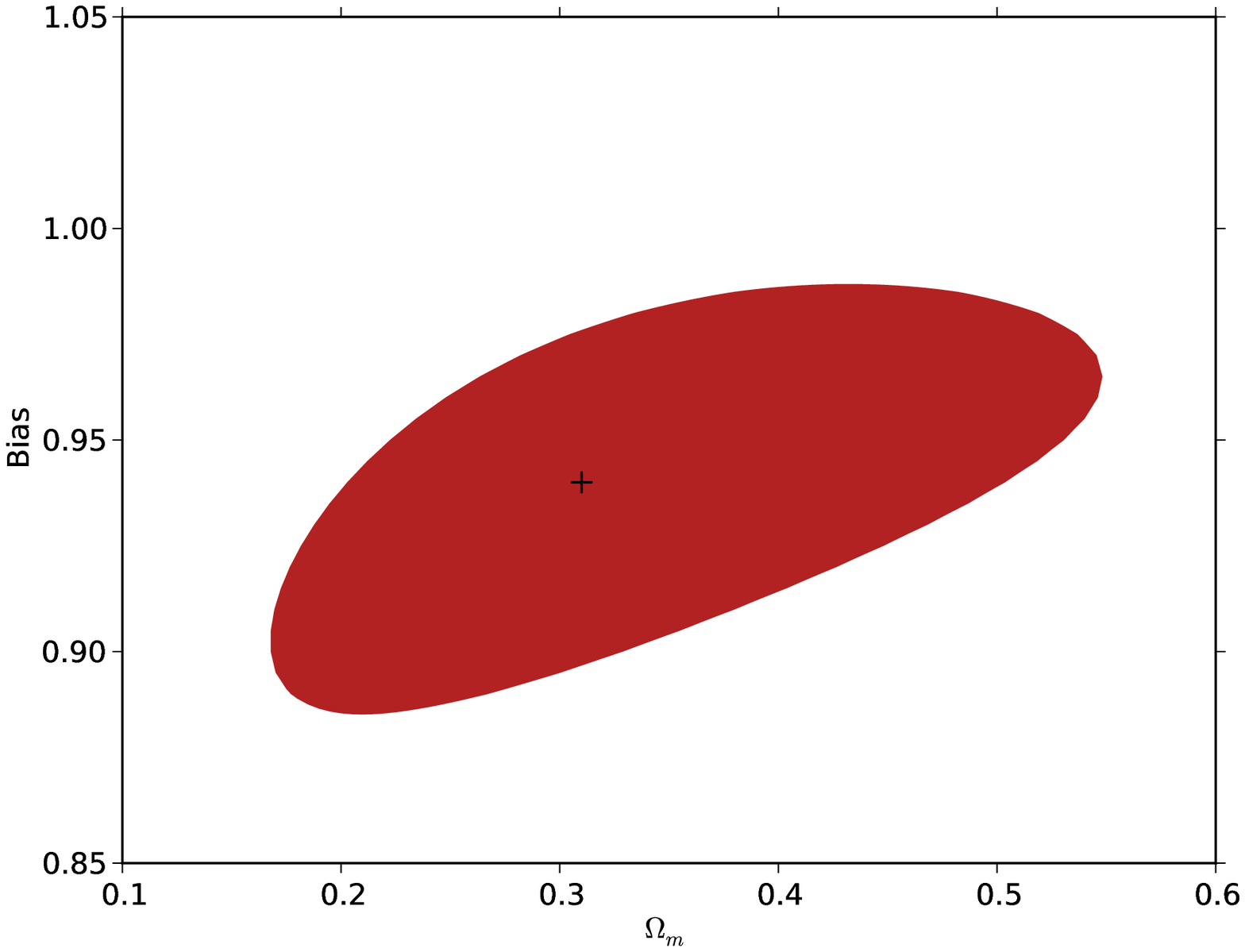}
    \caption{The black point at $\Omega_m = 0.31$, $b = 0.94$ is the minimum of the $\chi^2$ test for the entire sample, the area in red covers the 68\% confidence level.}
    \label{Chi Square 1}
  \end{center}
\end{figure}

We assume a flat cosmology and the WMAP baryon to matter ratio of $\Omega_b / \Omega_m = 0.168$ \citep{larson11} to perform this $\chi^2$ minimization 
for $\ell < 90$.  Over this range, the equivalent k is less than 0.16 h/Mpc at our median redshift of $\sim 0.2$; and, we therefore expect the linear $P(k)$ 
to be a good approximation.  We note that, given the limited range of the data used with this cut, the $\ell < 90$ restriction is not likely to yield 
competitive constraints on $\Omega_m$, and to fit the data past $\ell = 90$ we would need to use a non-linear power spectrum.  Indeed, we find a wide range of allowed $\Omega_m$ values, 
which we illustrate by displaying the results of our $\chi^2$ minimization for the 18--21 magnitude sample in Figure \ref{Chi Square 1}.
We find our best fit $\Omega_m = 0.31^{+0.18}_{-0.11}$ and $b = 0.94 \pm 0.04$ for the 18--21 sample, $\Omega_m = 0.26^{+0.25}_{-0.15}$ 
and $b = 1.09 \pm 0.05$ for the 18--19 magnitude subsample, $\Omega_m = 0.26^{+0.17}_{-0.11}$ and $b = 1.03 \pm 0.04$ for the 19--20 magnitude subsample, 
and $\Omega_m = 0.33^{+0.17}_{-0.10}$ and $b = 0.92 \pm 0.04$ 20--21 magnitude subsample.  We display these best-fit models against our measurements in 
Figure \ref{APS 2}.

\section{Discussion}
\label{Discussion}

Comparing the observed angular power spectrum to a linear theoretical spectrum is not expected to provide strong constraints
on $\Omega_m$ since varying $\Omega_m$ primarily changes the angular power spectrum at higher $\ell$.  Though weakly constrained, 
these measurements of $\Omega_m$ are consistent with other recent measurements of $\Omega_m$ from galaxy angular power spectra 
such as \cite{huterer01,frith05,blake07,thomas10}, as well as measurements through other methods such as the 7-year WMAP results 
from the cosmic microwave background \citep{larson11}.  This agreement confirms that the samples of galaxies and the measurement 
techniques we use have no large systematic errors.  

If we assume that the primordial fluctuations that seeded the large scale structure that we see today were Gaussian (e.g., \citealt{guth81}),
the angular power spectrum contains all clustering information on linear scales.  However, there has been some evidence that 
this might not be the case (e.g., \citealt{elsner10}).  Furthermore, non-linear effects from gravitational collapse become more 
pronounced at higher $\ell$, which also causes a departure from Gaussianity.  Though the quadratic estimator we employ assumes Gaussian fluctuations, 
the maximum likelihood angular power spectrum values we determine are unaffected by potential non-Gaussianities in the galaxy density field.  We note, 
however, that the presence of such non-Gausianities would generally cause us to underestimate our error bars (T02).

As we estimate the mean galaxy density from the survey itself, we constrain the data vector $\textbf{x}$ to have zero mean;
this is known as the integral constraint (see \citealt{tegmark98b} for a detailed discussion).  If we fail to account for the 
integral constraint we can underestimate the power on large scales \citep{huterer01}, so we correct for this by adding a large
number $M$ to the mean mode in the noise matrix $\textbf{N}$ before KL-compression.  The KL-compression stage will
determine that the signal-to-noise of the mean mode is low and it will be discarded with other low signal-to-noise modes.

The major limitation of our adopted approach for the calculation of the galaxy angular power spectrum is the computational 
difficulty.  The signal-to-noise of the SDSS DR7 is sufficient to
calculate the angular power spectrum to smaller scales than we have here, but doubling the resolution quadruples the number of 
pixels to $n_p \approx 25,000$.  Since the matrix muliplication and inversion scales as $O(n^3)$, doubling the resolution is a 
64-fold increase in computation, which is beyond our current computational resources, though we are looking into the 
possibility of performing this calculation, perhaps by KL-compressing the data even further.  

We have also explored using alternative platforms to accelerate the computation.  We have implemented this method on 
Graphics Processing Units (GPUs), which are part of every modern personal computer. GPUs 
are specifically designed to parallelize simple computations across many small multiprocessors, which make it ideal for 
vector and matrix calculations.  Using an Nvidia 8800 GTX and transferring the matrix operations to the GPU, while 
the rest of the code ran on the CPU, proved to be very effective at accelerating the quadratic estimation section of this 
calculation, speeding it up by a factor of 337.  For this to be effective, however, the matrices had to fit into the relatively 
small on board memory of the GPU, which in our test system was 768 MB.  In comparison, the memory required by the calculation 
performed in this paper was roughly 75 GB.  So while this platform seems very promising in accelerating this computation, the 
memory available will not be sufficient in the near future to allow us to meet or exceed the calculations that can be performed 
using current supercomputers.

While important, this work has merely been the first step.  By applying this method to volume-limited samples, we can constrain
the redshift evolution of the galaxy angular power spectrum.  In addition, we can use the photometric galaxy type classification
to distinguish differences in the clustering properties of early- and late-type galaxies in different redshift shells.  
Furthermore, by utilizing a full 3D, nonlinear theoretical power spectrum, we can model our measurements to higher $\ell$ values
and make more stringent measurements of cosmological parameters and we plan on taking these steps in a future work.

\section{Conclusions} 
\label{Conclusion}

We have used the quadratic estimation method with KL-compression to determine the SDSS DR7 angular power spectrum, first as a means of 
radical compression of the angular clustering information, and second to match these observed angular power spectra with theoretical 
angular power spectra to extract the linear bias and cosmological matter density.  We masked for observational effects and applied this
method to over 18 million SDSS DR7 galaxies and three magnitude subsamples out to $\ell \le 200$.  We also measured the angular power 
spectrum for each individual stripe out to $\ell \le 1000$ for stripes 9--37.  We have used the photometric redshift distribution of 
these galaxies to project the 3D power spectrum to two dimensions to obtain theoretical linear angular power spectrum, and used $\chi^2$ 
minimization to determine the best fit parameters given the observations.  As the linear angular power spectrum approximation is not valid
for the entire range of our estimated angular power spectrum, these parameter constraints have a large allowed range of values.

We found that the linear bias of our samples was $b = 1.09 \pm 0.05$ in the 18--19 magnitude range, $b = 1.03 \pm 0.04$ for 19--20, 
and $b = 0.92 \pm 0.04$ for 20--21, with an overall bias of $b = 0.94 \pm 0.04$ for our combined 18--21 magnitude sample.  We have 
also calculated the cosmological density of matter as $\Omega_m = 0.31^{+0.18}_{-0.11}$ from our entire sample.  

\section*{Acknowledgements}
\label{Acknowledgements}

The authors would like to thank Adam Myers, Dragan Huterer, Lloyd Knox for their valuable discussion and advice, and the referee for suggestions 
that have greatly improved the manuscript.

This research was supported in part by the National Science Foundation through XSEDE resources provided by Pittsburgh Supercomputing Center's 4,096 
core SGI UV 1000 (Blacklight) and 768 core SGI Altix (Pople) as well as the 1,024 processor SGI Altix (Cobalt) and 1,536 processor SGI Altix (Ember) 
of the National Center for Supercomputing Applications.

Funding for the SDSS and SDSS-II has been provided by the Alfred P. Sloan Foundation, the Participating Institutions, the National Science Foundation, 
the U.S. Department of Energy, the National Aeronautics and Space Administration, the Japanese Monbukagakusho, the Max Planck Society, and the Higher 
Education Funding Council for England. The SDSS Web Site is http://www.sdss.org/.

The SDSS is managed by the Astrophysical Research Consortium for the Participating Institutions. The Participating Institutions are the American Museum 
of Natural History, Astrophysical Institute Potsdam, University of Basel, University of Cambridge, Case Western Reserve University, University of Chicago, 
Drexel University, Fermilab, the Institute for Advanced Study, the Japan Participation Group, Johns Hopkins University, the Joint Institute for Nuclear 
Astrophysics, the Kavli Institute for Particle Astrophysics and Cosmology, the Korean Scientist Group, the Chinese Academy of Sciences (LAMOST), Los 
Alamos National Laboratory, the Max-Planck-Institute for Astronomy (MPIA), the Max-Planck-Institute for Astrophysics (MPA), New Mexico State University, 
Ohio State University, University of Pittsburgh, University of Portsmouth, Princeton University, the United States Naval Observatory, and the University 
of Washington.

Some of the results in this paper have been derived using the HEALPix \citep{gorski05} package.

\end{document}